\documentclass[twocolumn]{aastex631}

\usepackage{amsmath}
\usepackage{bm}
\usepackage{graphicx}
\usepackage{booktabs}
\usepackage{multirow}
\usepackage{xcolor}
\usepackage[caption=false]{subfig}

\hypersetup{
    colorlinks=true,
    linkcolor=blue,
    citecolor=blue,
    urlcolor=blue 
}

\newcommand{\inp}[2]{\left\langle #1 \middle| #2 \right\rangle}

\begin{document}

\title{On the Presence of a Tertiary Compact Object in GW190814}

\author[0000-0002-9523-7945]{Lalit Pathak}
\email{lalit.pathak@iucaa.in}
\affiliation{Inter-University Centre for Astronomy and Astrophysics, Post Bag 4, Ganeshkhind, Pune 411007, India}

\author[0000-0002-0471-3724]{Hemantakumar Phurailatpam}
\email{hemantakumar@phy.cuhk.edu.hk}
\affiliation{Department of Physics, The Chinese University of
Hong Kong, Shatin, New Territories, Hong Kong}

\author[0000-0003-4274-4369]{Achamveedu Gopakumar}
\email{gopu@tifr.res.in}
\affiliation{Department of Astronomy \& Astrophysics, Tata Institute of Fundamental Research, 1, Homi Bhabha Road, Mumbai- 400005, Maharashtra, India.}

\begin{abstract}

Gravitational waves from merging compact binaries are sensitive to line-of-sight acceleration (LOSA) induced by a massive companion in their vicinity. 
Interestingly, the leading-order contributions of LOSA and residual orbital eccentricity to the Fourier phase of the inspiral waveform exhibit similar frequency dependence, raising the possibility that a small eccentricity could mimic LOSA effects in transient GW events such as GW190814. We perform Bayesian inference using the \texttt{IMRPhenomXPHM} waveform family as the baseline LIGO–Virgo–KAGRA waveform model, augmented with leading-order LOSA and residual eccentricity corrections 
while using 32 seconds
of data associated with GW190814.
For a LOSA-only analysis, we find no evidence for a non-zero LOSA effect in GW190814, with a Bayes factor relative to the baseline model of $\sim 0.22$, consistent with the findings of  ~\cite{hendriks2026_LOSA} and in tension with the claim by \cite{Yang2025_LOSA}
who employed only 4 seconds of GW190814 data.
In a joint analysis that includes both leading order LOSA and eccentricity effects, we obtain informative posteriors for both parameters, with representative values $a/c \sim -2.8 \times 10^{-3} \,\mathrm{s}^{-1}$ and $e_0 \sim 0.11$.
However, the corresponding Bayes factor relative to the baseline model is $\sim 0.64$, suggesting that the 32-second data do not provide significant evidence for either LOSA or residual eccentricity in GW190814.
Further, our Bayesian runs reveal a strong correlation between the LOSA and eccentricity parameters, indicating a significant degeneracy in their imprint on the inspiral phase. This finding is consistent with our theoretical arguments and most likely explains the non-zero parameter estimates obtained in the joint analysis.
\end{abstract}

\section{Introduction}
The LIGO-Virgo-KAGRA (LVK) network of gravitational-wave (GW) observatories \citep{advLIGO_2015,Acernese_2014,Abbott_2020, akutsu2020overviewkagradetectordesign, PhysRevD.88.043007} has inaugurated the era of GW astronomy by detecting transient GWs from merging black holes and neutron stars \citep{PhysRevLett.116.061102,LIGOScientific:2017vwq}. As of the latest publicly released Gravitational-Wave Transient Catalog (GWTC-4.0) \citep{LIGOScientific:2025hdt,LIGOScientific:2025slb,LIGOScientific:2025yae}, the LVK collaboration has reported $\sim 218$ such events across a wide range of masses. Subsequent observing runs (O4b and O4c), which have been completed but are not yet publicly released, are expected to add hundreds more. 
In the 2030s, LIGO-India  is expected to join the existing network of ground-based GW detectors operating near their design sensitivities
~\citep{Saleem_2021,Shukla:2023kuj}.
This should enable the LVK collaboration to approach the full potential of second-generation GW ~\citep{advLIGO_2015,Acernese_2014,Abbott_2020}.
The majority of detections to date correspond to binary black holes (BBHs), primarily due to selection effects in ground-based GW detectors \citep{Chen_2017, Gerosa_2024}.

With the growing number of detections, we are beginning to probe the underlying population of BBHs, their formation channels \citep{MANDEL20221}, and their evolutionary pathways \citep{LIGOScientific:2025pvj, PhysRevX.13.011048}. BBHs are expected to form either through isolated binary stellar evolution \citep{Dominik_2012, Dominik_2013, Dominik_2015, Belczynski_2016, Silsbee_2017, Murguia_Berthier_2017, Rodriguez_2018, Schr_der_2018, Iorio_2023} or via dynamical interactions in dense stellar environments \citep{Portegies_Zwart_2000, Lee_2010, Banerjee_2009, Tanikawa_2013, Bae_2014, PhysRevLett.115.051101, Ramirez_Ruiz_2015, PhysRevD.93.084029, Rodriguez_2016, Askar_2016, Park_2017, PhysRevD.97.103014, Samsing_2018, Samsing_2020, Trani_2019, Liu_2021, Atallah_2023} such as star clusters \citep{Abbott_2016, PhysRevX.9.031040, Abbott_2020}. In addition, active galactic nucleus (AGN) disks \citep{McKernan_2012, Peng_2021, tagawa2026electromagneticflarescompactobjectmergers, Tagawa_2020, Bartos:2016dgn, McKernan_2018, Stone_2016, Samsing_2022, trani2024threebodyencountersblackhole, Fabj_2024} have been proposed as a potential site for BBH formation and mergers.

The populations of BBHs arising from different formation channels are expected to exhibit distinct signatures in their parameter distributions, including masses \citep{Zevin_2017, Su_2021}, spins \citep{Kalogera_2000, Rodriguez_2016, Liu_2018}, and orbital eccentricities \citep{Zevin_2019, PhysRevD.100.043010, Samsing_2020, Gultekin_2004}. In recent years, increasing attention has been directed towards probing environmental effects (EEs) as potential indicators of specific formation scenarios. These effects can introduce additional phase corrections on top of the vacuum general relativity (GR) waveform of a binary. Such dephasing contributions are typically parameterized by additional degrees of freedom that can be constrained by GW observations. The inferred distributions of these parameters can then be used to either place stringent constraints on, or potentially rule out, the environments in which these BBHs form and merge \citep{zwick2025environmentaleffectsstellarmass}.

One of the prominent environmental effects that has recently gained attention is the so-called line-of-sight acceleration (LOSA) \citep{Meiron_2017,PhysRevD.95.044029,Vijaykumar_2023,samsing2024gravitationalwavephaseshifts,PhysRevD.98.064012}. LOSA arises when a binary is subject to the gravitational potential of a third body, leading to a secular modulation of the GW phase. At leading order, this effect exhibits a characteristic frequency dependence scaling as $f^{-13/3}$, imprinting a distinct signature on the waveform that can, in principle, be measured by GW detectors. \cite{Yang2025_LOSA} recently reported a non-zero measurement of LOSA ($a/c \sim  1.5 \times 10^{-3} \mathrm{s}^{-1}$,
where $a$ specifies LOSA and $c$ stands for the speed of light
) in GW190814~\citep{GW190814LVK}, claiming a strong statistical preference over an isolated binary scenario. However, \cite{hendriks2026_LOSA} reanalyzed the same event using Bayesian inference with a more conservative signal duration of $32$ seconds and found no evidence for LOSA, attributing the earlier claim to the use of a shorter signal duration of $4$ seconds, which they argue can lead to biased or spurious inferences.

We demonstrate that the dominant contributions of LOSA and residual orbital eccentricity to the Fourier phase of the inspiral waveform depend on the GW frequency in a similar manner. Specifically, the LOSA and residual orbital eccentricity effects modify the Fourier phase of the inspiral waveform with characteristic scalings of $f^{-13/3}$ and $f^{-34/9}$, respectively~\citep{Vijaykumar_2023,hemanta2025eccentricity}. This similar frequency dependence in the waveform phase evolution implies that their effects may become partially degenerate in the observed signal, potentially leading to biased inference if only one of these effects is modelled. Indeed, we show that GW phase modulation attributed to LOSA can instead be mimicked by a small residual eccentricity. \cite{tiwari2025b} also discussed orbital eccentricity as a potential mimicker of LOSA. Furthermore,~\cite{eect_sajad} developed the Eccentricity Evolution Consistency Test (EECT) for distinguishing different mimickers of eccentricity, including LOSA, and discussed the LOSA--eccentricity degeneracy in the context of eccentricity mimickers (see Sections III.C and V.B.2). In the present work, we explicitly highlight that the LOSA--eccentricity degeneracy originates from the similar frequency scaling of their leading-order phase corrections, and explore its implications for Bayesian inference and parameter estimation of transient GW events such as GW190814.

We then perform standard Bayesian PE studies of GW190814 to probe the extent to which these two effects can be disentangled using GW observations and to substantiate recent inferences on this event \citep{Yang2025_LOSA,hendriks2026_LOSA}.
To construct our Inspiral-Merger-Ringdown (IMR) waveform model that incorporates the effects of both LOSA and residual eccentricity, we modify the phase of the widely used \texttt{IMRPhenomXPHM} approximant~\citep{PhysRevD.103.104056} by introducing appropriate Fourier phase corrections for these two effects.

In our LOSA-only analysis using a $32$ seconds signal duration for GW190814, we find no evidence for a non-zero LOSA, with a Bayes factor (BF) of $\sim 0.22$ relative to the baseline model. In contrast, repeating the analysis with a $4$ seconds signal duration yields a non-zero estimate of LOSA with high statistical significance, supporting the conclusion of \cite{hendriks2026_LOSA} that shorter durations can lead to biased inferences.
In the joint analysis including both LOSA and eccentricity, we find a strong correlation between the two parameters, with representative values $a/c \sim  -2.8 \times 10^{-3} \mathrm{s}^{-1}$ and $e_0 \sim 0.11$. This is consistent with the expected degeneracy arising from their similar frequency dependence. However, the corresponding Bayes factor is $\sim 0.64$ relative to the baseline model without these effects, indicating that the data do not provide significant evidence for either LOSA or eccentricity.

For the eccentricity-only analysis with a $32$ seconds signal duration, we find no evidence for non-zero eccentricity, consistent with previous Bayesian studies of GW190814 \citep{Kacanja_2025}. Additionally, restricting the joint analysis to the dominant $(2,2)$ mode of the \texttt{IMRPhenomXPHM} approximant yields essentially uninformative (flat) posteriors for both $a/c$ and $e_0$ (for both individual and joint analysis), highlighting the importance of higher-order modes (HOMs), even if their contribution to the signal-to-noise ratio (SNR) is modest. We discuss various implications of our results for upcoming observational runs and detector upgrades.

The rest of the paper is organized as follows. In Sec.~\ref{sec:waveform_model}, we describe the waveform model used in the analysis and introduce the phase corrections arising from LOSA and eccentricity. We also compute the match between a quasi-circular BBH waveform and a waveform including both LOSA and eccentricity. In Sec.~\ref{subsec:bayesian_inference_basics}, we introduce the basics of PE and briefly review the Bayesian framework for inferring GW parameters from detector strain data.  In Sec .~\ref {subsec:losa_only_analysis}, \ref{subsec:e0_only_analysis}, and \ref{subsec:losa_e0_joint_analysis}, we present the results of the various PE analyses performed for GW190814. Finally, in Sec.~\ref{sec:conclusion}, we summarize our findings and discuss future directions.

\section{Modelling LOSA and
\lowercase{$e_0$}
Effects}
\label{sec:waveform_model}
 We begin by explaining how we adapt and modify an existing LVK-relevant Inspiral-Merger-Ringdown (IMR) waveform family to explore the combined effects of LOSA and residual orbital eccentricity in merging compact binaries (MCBs). This section also discusses the preliminary data analysis implications of such a modified IMR waveform family.

For our main analysis, we employ a modified version of \texttt{IMRPhenomXPHM}, a frequency-domain phenomenological waveform model widely used by the LVK collaboration for PE of detected GW events~\citep{PhysRevD.103.104056}. This model describes the inspiral, merger, and ringdown of precessing binary black holes and accounts for HOMs. The inspiral part of \texttt{IMRPhenomXPHM} is built upon the \texttt{TaylorF2}~\citep{PhysRevD.49.1707, PhysRevD.59.124016, Buonanno_2009, Faye_2012, PhysRevLett.74.3515} approximant, which is derived by applying the stationary phase approximation to the time-domain post-Newtonian (PN) inspiral waveform, yielding an analytic expression for the Fourier phase. The \texttt{TaylorF2} approximant, present in 
\texttt{IMRPhenomXPHM}, incorporates PN corrections up to 3.5PN order for the phase and 2PN order for the amplitude. This PN inspiral is then seamlessly stitched to phenomenological merger and ringdown waveforms calibrated to numerical relativity (NR) simulations~\citep{Boyle_2019}.
We note in passing that we employed 
\texttt{IMRPhenomXPHM}, 
as implemented in \texttt{bilby}~\citep{Ashton_2019, Romero_Shaw_2020},
 as our baseline waveform model. Further, we invoked 
 the \texttt{SpinTaylor}~\citep{colleoni2024fastfrequencydomaingravitationalwaveforms} extension for \texttt{IMRPhenomXPHM} which was used for LVK's GWTC-4.0~\citep{gwtc4} catalogue, influenced by 
  ~\cite{hendriks2026_LOSA}.
  
  Note that the post-Newtonian (PN) approximation provides a perturbative framework for modelling the inspiral dynamics of compact binaries in the weak-field, slow-velocity regime, expanding the 
 orbital dynamics and GW  flux in powers of $(v/c)^2$, where $v$ is the binary orbital speed. For the present work, we are primarily interested in modifying 
 \texttt{IMRPhenomXPHM} to incorporate the effects of LOSA and residual eccentricity. Specifically, we introduce additional Fourier phase corrections that capture the leading-order contributions of these two effects, whose similar frequency dependence ($f^{-13/3}$ for LOSA and $f^{-34/9}$ for eccentricity) motivates our investigation of their potential degeneracy in observed signals.

If the center of mass of an MCB experiences a non-zero, time-varying velocity component along the line of sight- for instance, if it orbits a supermassive black hole- the higher time derivatives of the orbital velocity can contribute to the GW Fourier phase, as demonstrated in~\citep{Vijaykumar_2023}. This is because the detector-frame total mass $M_{\rm det}$ of such an MCB becomes time dependent:
\begin{equation}
M_{\rm det} = M_{\rm src} \, (1 + z_{\rm cos}) \, (1 + z_{\rm dop}) \, \left(1 + \frac{a}{c} \times t \right),
\end{equation}
where $M_{\rm src}$ is the source-frame total mass. This equation accounts for the effects of cosmological redshift $z_{\rm cos}$, Doppler redshift $z_{\rm dop} = v/c$ due to a constant line-of-sight velocity, and the time-dependent LOSA effect via the $(a/c)\, t$ term. Here we assume $z_{\rm dop} \ll 1$ and $|a/c| \times t \ll 1$, consistent with a non-relativistic, slowly varying line-of-sight motion.

This time-dependent mass introduces an additional correction to the Fourier phase of the \texttt{TaylorF2} approximant. For the dominant $(2,2)$ mode at leading order, the phase correction due to LOSA is given by \citep{PhysRevD.83.044030,PhysRevD.95.044029, PhysRevD.101.063002, Vijaykumar_2023}:

\begin{equation}
    \begin{aligned}
        \Delta \Psi_{\rm LOSA} &= \frac{25}{65536\, \eta^2} \left(\frac{GM}{c^3}\right)\left(\frac{a}{c}\right)\, \nu_f^{-13} \\
        &= \frac{25\, \pi^{-13/3}}{65536\, \eta^2} \left(\frac{GM}{c^3}\right)^{-10/3}\left(\frac{a}{c}\right)\, f^{-13/3}
    \end{aligned}
    \label{eq:phase_lsa}
\end{equation}
where $\nu_f = (\pi GM f / c^3)^{1/3}$ is the dimensionless frequency parameter, $M$ is the detector-frame total mass with component masses $m_1$ and $m_2\,$,
and $\eta = m_1 m_2 / (m_1 + m_2)^2$ is the symmetric mass ratio. The scaling
$\Delta \Psi_{\rm LOSA} \propto f^{-13/3}$ is characteristic of the LOSA effect and serves as a key signature distinguishing it from other environmental or dynamical effects.

We now incorporate the effects of residual orbital eccentricity into the Fourier-domain phase of the waveform. The corresponding leading-order phase correction is given by:
\begin{equation}
\begin{aligned}
\Delta \Psi_{e_0} =\,& \frac{6}{256\eta} 
\left(\frac{2\pi GM}{c^3}\right)^{-5/3}
f_{\rm low}^{19/9} f^{-34/9} \\
&\times \Bigg[
\frac{2355\,e_0^2}{1462} \\
&\quad - \left(
\frac{5222765\, f_{\rm low}^{19/9} f^{11/9}}{998944}
- \frac{2608555}{444448}
\right)e_0^4
\Bigg] \,,
\label{eq:phase_eccentric_2_2_additional_new_e0_final}
\end{aligned}
\end{equation}
where  $ f_{\mathrm{low}} $  denotes the lower cutoff frequency at which the eccentricity is defined as $ e_0 $, and we let $f_{\mathrm{low}}$ to be $20\,$Hz in our PE runs.
It is evident from the expression that the leading-order contribution, proportional to  $ e_0^2 $, scales as  $  f^{-34/9}$. 
 We would like to note that the above expression is adapted from the \texttt{TaylorF2Ecck}~\citep{hemanta2025eccentricity} approximant, which is compatible with the LVK Algorithm Library Suite
 \citep{lalsuite}.
 This approximant models inspiral GWs from non-spinning compact binaries on PN-accurate eccentric orbits, while restricting initial-eccentricity contributions to their leading order for computational tractability. It is a fully analytic, frequency-domain inspiral model that consistently incorporates three key physical effects up to 3PN order: orbital dynamics, advance of periastron, and GW emission and incorporates 
 inputs from~\cite{Damour_2004, Tiwari_2019, Tanay_2016, Boetzel_2017}.

To perform Bayesian PE studies probing the effects of LOSA and $ e_0 $, both individually and jointly, we modify the Fourier-domain phase of \texttt{IMRPhenomXPHM} waveform model, which provides a complete inspiral-merger-ringdown description for precessing, eccentricity-free binaries. Influenced by~\cite{hendriks2026_LOSA}, our modification takes the following form:
\begin{equation}
\tilde{h}_{\rm tot}(f) = \tilde{h}_{\rm base}(f) \; \exp\left[-i \, \Delta \Psi_{\rm deph}(f)\right]
\label{eq:htot}
\end{equation}

where $ \tilde{h}_{\rm base}(f) $ is the frequency-domain  IMR waveform from the standard (non-eccentric, non-LOSA) \texttt{IMRPhenomXPHM}  model, and 
$ \Delta \Psi_{\rm deph}(f) $  is an {\it additional dephasing} chosen according to the physical effect under investigation.
In other words, we let 
\begin{equation}
\Delta \Psi_{\rm deph}(f) =
\begin{cases}
\Delta \Psi_{\rm LOSA}(f) & \text{(LOSA only)}\\
\Delta \Psi_{e_0}(f) & \text{( $e_0$  only)}\\
\Delta \Psi_{\rm both}(f) & \text{(both LOSA and $ e_0$ )}
\end{cases}
\label{eq:dephasing_equation}
\end{equation}
where $\Delta \Psi_{\rm both}(f) = \Delta \Psi_{\rm LOSA}(f) + \Delta \Psi_{e_0}(f)$. 
Note that this multiplicative phase-factor approach ensures modularity: the baseline waveform remains unchanged, and physical extensions are cleanly isolated in the phase domain, facilitating efficient and interpretable Bayesian inference.

\begin{figure}
    \centering
    \includegraphics[width=\linewidth]{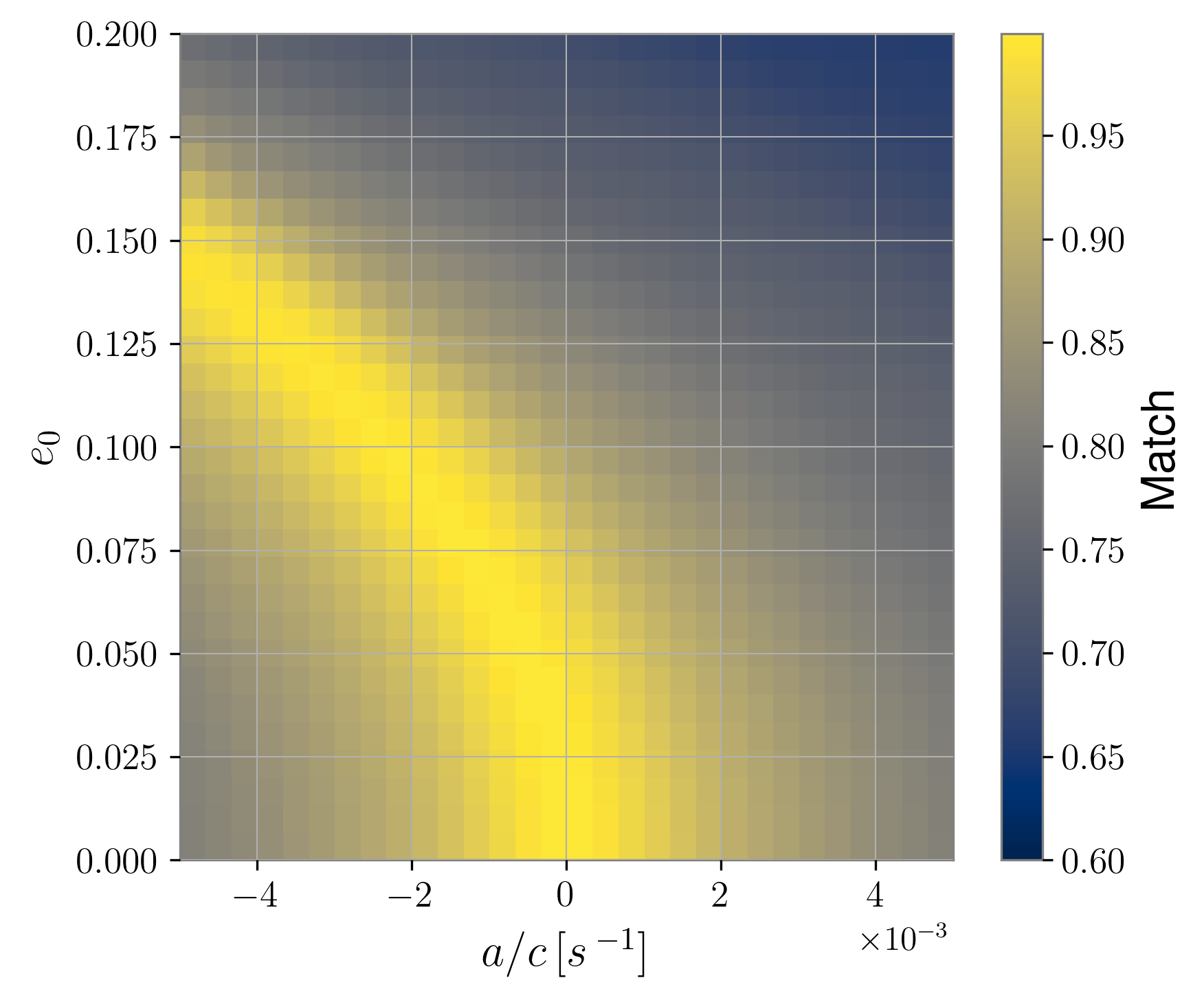}
    \caption{
We plot $\mathcal{M}(h_1,h_2)$ as a function of the LOSA parameter $a/c$ and eccentricity $e_0$ for a GW190814-like signal. The two waveform families involved are the traditional \texttt{IMRPhenomXPHM} waveform and its modified version that incorporates the dominant-order effects of LOSA and $e_0$. Values of match closer to unity indicate higher similarity between waveforms. The diagonal ridge of high match values reveals a strong correlation between $a/c$ and $e_0$, demonstrating a degeneracy in their imprint on the GW phase, whereby different combinations of LOSA and eccentricity produce nearly indistinguishable waveforms ($\rm match > 0.97$). The displayed match values are computed using the LIGO-Hanford (H1) detector PSD, maximizing over time and phase shifts. 
    }
    \label{fig:match_a_LSA_e0_H1}
\end{figure}

A few comments are in order. When invoking these modifications in our Bayesian PE runs, we mainly focus on the leading-order LOSA corrections, as evident from Eq.~\ref{eq:phase_lsa}. This choice is motivated by the fact that the signal-to-noise ratio (SNR) contributions from higher-order terms are not sufficiently large to significantly impact our conclusions. Furthermore, we consider only the leading-order contribution in eccentricity (see Eq.~\ref{eq:phase_eccentric_2_2_additional_new_e0_final}), as the higher-order 
$ e_0 $ contributions were found to have a negligible effect on the results. Nevertheless, for completeness, we perform an additional analysis that includes higher-order eccentricity corrections up to  $ \mathcal{O}(e_0^4) $, and find no appreciable difference in our main conclusions. 
To probe the preliminary data analysis implications of our $\tilde{h}_{\rm tot}(f)$ model, we perform match estimates employing both the $\tilde{h}_{\rm tot}(f)$ and $\tilde{h}_{\rm base}(f)$ waveform families, as defined in Eq.~\ref{eq:htot}.
Recall that the match between two waveform families $h_1$ and $h_2$ quantifies their similarity. It is defined as the noise-weighted inner product between normalized waveforms, maximized over an overall phase ($\varphi_{\rm ref}$) and time shift ($t_{\rm ref}$), which are kinematical variables present in the waveform families. The resulting match is given by~\citep{findchirp_b_allen, k21q-wp8f}:
\begin{equation}
    \mathcal{M}(h_1, h_2) = \max_{t_{\rm ref}, \varphi_{\rm ref}} \: \inp{\hat{h}_1}{\hat{h}_2},
    \label{eq:Match}
\end{equation}
where $\hat{h}_1$ denotes the normalized waveform, defined as $\hat{h}_1 = h_1 / \sqrt{\inp{h_1}{h_1}}$, and $\inp{h_1}{h_2}$ is the noise-weighted inner product between two waveforms $h_1(t)$ and $h_2(t)$, given by
\begin{equation}
    \inp{h_1}{h_2} = 4 \, \Re \int_{f_{\rm low}}^{f_{\rm high}} 
    \frac{\tilde{h}_1^{*}(f)\,\tilde{h}_2(f)}{S_n(f)} \, df,
    \label{eq:innerProduct}
\end{equation}
where $S_n(f)$ denotes the one-sided noise power spectral density (PSD), and $\tilde{h}_1(f)$ is the Fourier transform of $h_1(t)$. For the present effort, we use the maximum-likelihood sample and PSD from the publicly available GW190814 posterior samples provided by the LVK Collaboration~\citep{GW190814_posterior}.

As noted earlier, this analysis examines whether there is a strong correlation among the free parameters in our IMR family, which incorporates both LOSA and $e_0$ effects, compared to the conventional IMR family that ignores these additional astrophysical possibilities. Theoretically, we expect a degree of correlation between these two effects, since at leading order they exhibit a similar frequency dependence (scaling approximately as $\sim f^{-4}$).
In Fig.~\ref{fig:match_a_LSA_e0_H1}, we display \( \mathcal{M}(\tilde{h}_{\rm tot}(f), \tilde{h}_{\rm base}(f) ) \) as a function of \(e_0\) and \(a/c\). The figure clearly shows a strong correlation between LOSA (\(a/c\)) and eccentricity (\(e_0\)), especially for higher match values. This provides strong motivation to analyze the GW190814 event using \(\tilde{h}_{\rm tot}(f)\), rather than probing the presence of individual LOSA and \(e_0\) effects in isolation. We note in passing that we employed the H1 detector PSD for the match calculation; however, a similar structure is observed for other detector PSDs, with a slightly broader spread depending on their respective sensitivities\footnote{We also perform several zero-noise injection analyses for GW190814-like signals to assess the recovery of LOSA and eccentricity parameters and to investigate the impact of HOMs on the inference. These studies further illustrate the strong LOSA-eccentricity degeneracy discussed in the main text. For more details, see Appendix~\ref{sec:appendix}.}.

\section{ Probing LOSA and 
\lowercase{$e_0$} effects in GW190814}
\label{sec:bayesian_inference}

We begin by providing a basic outline of the Bayesian inference approach used to perform PE studies of transient GW events, with particular focus on GW190814. Thereafter, we discuss our PE results for the various waveform cases described in Eq.~\ref{eq:dephasing_equation}, which yield IMR waveform families that incorporate LOSA and/or $e_0$ effects.

\subsection{Details of Bayesian Inference for GW190814}
\label{subsec:bayesian_inference_basics}
In GW astronomy, Bayesian inference is the standard framework for parameter estimation, enabling us to infer source properties, such as masses, spins, and eccentricity, from detector data. The method yields posterior probability distributions that quantify our knowledge of these parameters given the observed signal and our prior beliefs~\citep{PhysRevD.46.5236, PhysRevD.49.2658, LIGOScientific:2025hdt}.
Naturally, the approach requires 
 a GW template family $h(\vec{\Lambda})$, strain data $d$ from a ground-based GW detector, and a hypothesis/model $\mathcal{H}$, where $\vec{\Lambda}$ denotes the set of compact binary parameters (such as component masses, spins, orientation, and sky location). 
 With these inputs, the posterior distribution ${p(\vec{\Lambda} \mid d, \mathcal{H})}$ can be inferred using Bayes' theorem as
\begin{equation}
    p(\vec{\Lambda} \mid d, \mathcal{H}) = \frac{p(d \mid \vec{\Lambda}, \mathcal{H}) \, p(\vec{\Lambda} \mid \mathcal{H})}{p(d \mid \mathcal{H})}
    \label{eq:bayes_theorem}
\end{equation}

where ${p(\vec{\Lambda} \mid \mathcal{H})}$ denotes the prior distribution over the binary parameters, and ${p(d \mid \vec{\Lambda}, \mathcal{H})}$ is the GW likelihood~\citep{PhysRevD.91.042003, Thrane_Talbot_2019}, given by
\begin{equation}
    p(d \mid \vec{\Lambda}, \mathcal{H}) \propto \exp\left(-\frac{1}{2} \left\langle d - h(\vec{\Lambda}) \mid d - h(\vec{\Lambda}) \right\rangle \right)
\end{equation}

where $\langle \cdot \mid \cdot \rangle$ denotes the noise-weighted inner product as defined in Eq.~\ref{eq:innerProduct}.

The denominator in Eq.~\ref{eq:bayes_theorem}, ${p(d \mid \mathcal{H})}$, is the evidence (denoted by $\mathcal{Z}$), which is obtained by marginalizing the numerator over the parameter space of the model $\mathcal{H}$ as
\begin{equation}
    \mathcal{Z} = \int d\vec{\Lambda} \, p(d \mid \vec{\Lambda}, \mathcal{H}) \, p(\vec{\Lambda} \mid \mathcal{H})
    \label{eq:evidence}
\end{equation}

Using the evidence $\mathcal{Z}$, we can assess how well a model explains the observed data $d$ in comparison to another model. A commonly used measure for model comparison in Bayesian analysis is the \emph{Bayes factor} (BF)~\citep{Thrane_Talbot_2019}, defined as the ratio of the evidence for two competing models. A larger BF indicates stronger support for one hypothesis over the other. For two competing models $\mathcal{H}_1$ and $\mathcal{H}_2$, the BF is given by
\begin{equation}
    \mathrm{BF}^{\mathcal{H}_1}_{\mathcal{H}_2} = \frac{\mathcal{Z}_1}{\mathcal{Z}_2}
    \label{eq:bayes_factor}
\end{equation}

We can use $\Delta \, \log_{10} \rm BF^{\mathcal{H}_1}_{\mathcal{H}_2} \equiv \log_{10}\mathcal{Z}_1 - \log_{10}\mathcal{Z}_2$ as the distinguishability criterion for models, with values of $1-2$ indicating strong evidence and $>2$ indicating decisive distinguishability~\citep{hemanta2025eccentricity}. 
\begin{figure*}[!hbt]
    \centering
    \includegraphics[width=0.95\linewidth]{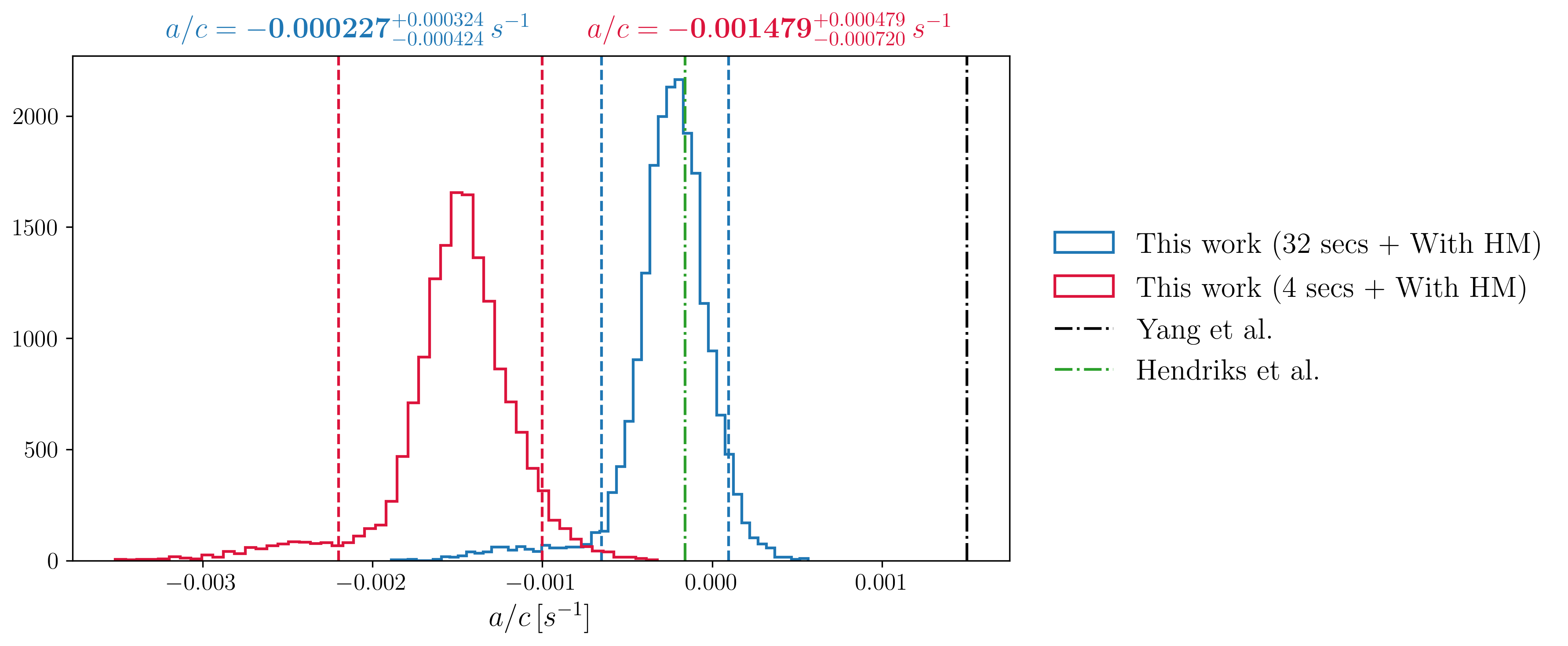}
    \caption{Posterior distributions of the LOSA parameter ($a/c$) for GW190814 obtained from LOSA-only analysis using different signal durations. The blue histogram shows results using a $32$ second data segment, while the red histogram corresponds to a $4$ seconds segment.
    Vertical dashed lines indicate the $90\%$ credible intervals for each case. The green dashed-dotted line marks the result reported by ~\cite{hendriks2026_LOSA}, while the black dash-dotted line denotes the value inferred by ~\cite{Yang2025_LOSA}. The $32$ seconds analysis yields a posterior consistent with zero LOSA, whereas the $4$ seconds analysis recovers a significantly non-zero value of $a/c$. This comparison illustrates the strong dependence of the LOSA inference on the chosen signal duration and highlights the potential for biased estimates when using shorter data segments.
    The plots are from our PE runs, in which we incorporate only the dominant-order LOSA effect into our baseline waveform model \texttt{IMRPhenomXPHM} that includes HOMs~\citep{PhysRevD.103.104056}.
    }
    \label{fig:losa_comparion_32sec_vs_4sec_withHM}
\end{figure*}

We perform detailed PE studies for the event, GW190814~\citep{GW190814LVK}  using publicly available strain data from the Gravitational Wave Open Science Centre (\texttt{GWOSC})~\citep{Abbott_2023}. Further, we adopt the same PSD used by the LVK Collaboration in their analysis of GW190814. 
For the standard MCB parameters, we use the same prior distributions as in ~\cite{Yang2025_LOSA} and ~\cite{hendriks2026_LOSA} (see Table~\ref{tab:prior_table}).
For analyses including LOSA, we assume a uniform prior on the acceleration parameter 
$a/c$  over the interval \([-0.005, 0.005]~\mathrm{s}^{-1}\). For analyses including eccentricity, we assume a uniform prior on the eccentricity parameter $e_0$  over the interval $[0, 0.2]$, encompassing the range of residual eccentricities expected for LVK sources without being overly restrictive~\citep{Kacanja_2025}.
To disentangle the effects of LOSA and eccentricity, we perform PE studies while including only one effect at a time and later jointly by including both effects simultaneously, as displayed in Eq.~\ref{eq:dephasing_equation}. For individual analyses, we fix the parameter corresponding to the other effect to zero via the prior (e.g., $e_0 = 0$ for LOSA-only analyses and $a/c = 0$ for $e_0$-only analyses). Additionally, we perform the PE studies for GW190814 for both with and without HOMs, excluding the effects of LOSA and eccentricity by setting $a/c = 0$ and $e=0$ in Eq.~\ref{eq:dephasing_equation} and in the priors (Table~\ref{tab:prior_table}). We refer to this configuration as the \emph{vanilla} (VAN) run throughout the paper. 

Our primary results use a signal duration of $32$ seconds, which is a conservative choice given the $\sim 10$ seconds inspiral duration of GW190814, as discussed in ~\cite{hendriks2026_LOSA}. Unless stated explicitly, we include all modes available in \texttt{IMRPhenomXPHM}, given the asymmetric nature of the GW190814 event. To sample the posterior distribution ${p(\vec{\Lambda} \mid d)}$, we use the nested sampling~\citep{skilling2006nested, higson2019dynamic} algorithm \texttt{dynesty}~\citep{speagle2020dynesty, sergey_koposov_2023_7600689}, as implemented in the \texttt{bilby}~\citep{Ashton_2019, Romero_Shaw_2020} software package. We use the following sampler settings: \texttt{nlive = 1024}, \texttt{dlogz = 0.1}, \texttt{sample = "acceptance-walk"}, and \texttt{naccept = 60}. 
 We now list results from our detailed PE runs. 
 
\subsection{LOSA-Only Analysis}
\label{subsec:losa_only_analysis}
For the LOSA-only analysis, we find no evidence for a non-zero LOSA, consistent with the recent non-detection reported by~\cite{hendriks2026_LOSA}. We also note that~\cite{tiwari2025b} argued that the reported LOSA signature may be attributable to waveform systematics.  
The Bayes factor comparing the LOSA-only run to the \emph{vanilla} run is $\mathrm{BF^{LOSA}_{VAN}} \sim 0.22$ corresponding to a $\Delta\log_{10} \mathrm{BF^{LOSA}_{VAN}} \sim -0.65$, indicating no support for the presence of LOSA in the GW190814 data. 
Therefore, our detailed PE runs suggest that the LOSA inference for GW190814 reported by ~\cite{Yang2025_LOSA} may be sensitive to the choice of signal duration, as their analysis used only 4 seconds of data, compared to our longer segment. 
For completeness, we also perform a subset of analyses using a $4$ seconds signal duration. In this case, we recover a non-zero estimate of LOSA in our LOSA-only analysis; however, the inferred value is negative\footnote{\cite{Yang2025_LOSA} reported the positive sign for $a/c$, which could potentially be related to a different sign convention adopted for the additional phase correction in their implementation.} $({a/c \sim  -1.4 \times 10^{-3} \mathrm{s}^{-1}})$ and a corresponding $\mathrm{BF^{LOSA}_{VAN}} \sim 567$ (and $\Delta\log_{10} \mathrm{BF^{LOSA}_{VAN}} \sim 2.75$). This result supports the argument by ~\cite{hendriks2026_LOSA} that the use of an insufficient signal duration may lead to spurious or unphysical inferences of LOSA. 

\begin{table}[t]
\centering
\small
\setlength{\tabcolsep}{5pt}

\caption{
\textbf{Prior parameter space} for the 17-dimensional parameter vector $\vec{\Lambda}$. For $\mathcal{M}$ and $q$, we use prior distributions such that it gives a uniform distribution in component masses. 
}
\label{tab:prior_table}

\begin{tabular}{lcc}
\toprule
Parameter & Range & Prior distribution \\
\midrule

$\mathcal{M}$ 
& $[6,\,7]\,M_\odot$ 
& Uniform \\

$q$ 
& $[0.1,\,1.0]$ 
& Uniform \\

$m_1$ 
& $[15,\,35]\,M_\odot$ 
& Constraint \\

$m_2$ 
& $[1,\,5]\,M_\odot$ 
& Constraint \\

$a_1$ 
& $[0,\,0.2]$ 
& Uniform \\

$a_2$ 
& $[0,\,1.0]$ 
& Uniform \\

$\theta_1$ 
& $[0,\,\pi]$ 
& Sine \\

$\theta_2$ 
& $[0,\,\pi]$ 
& Sine \\

$\Delta \phi$ 
& $[0,\,2\pi]$ 
& Uniform \\

$\phi_{\rm JL}$ 
& $[0,\,2\pi]$ 
& Uniform \\

$d_{\rm L}$ 
& $[167,\,280]\,\mathrm{Mpc}$ 
& PowerLaw$(\alpha=2)$ \\

$\delta$ 
& $[-\pi,\,\pi]$ 
& Cosine \\

$\alpha$ 
& $[0,\,2\pi]$ 
& Uniform \\

$\iota$ 
& $[0,\,\pi]$ 
& Sine \\

$\psi$ 
& $[0,\,\pi]$ 
& Uniform \\

$\phi$ 
& $[0,\,2\pi]$ 
& Uniform \\

$t_c$ 
& $[t_{\rm trig}-0.1,\,
t_{\rm trig}+0.1]\,\mathrm{s}$ 
& Uniform \\

$a/c$ 
& $[-0.005,\,0.005]\,\mathrm{s}^{-1}$ 
& Uniform \\

$e_0$ 
& $[0,\,0.2]$ 
& Uniform \\

\bottomrule
\end{tabular}
\end{table}

\begin{figure}[!hbt]
    \centering
    \includegraphics[width=\linewidth]{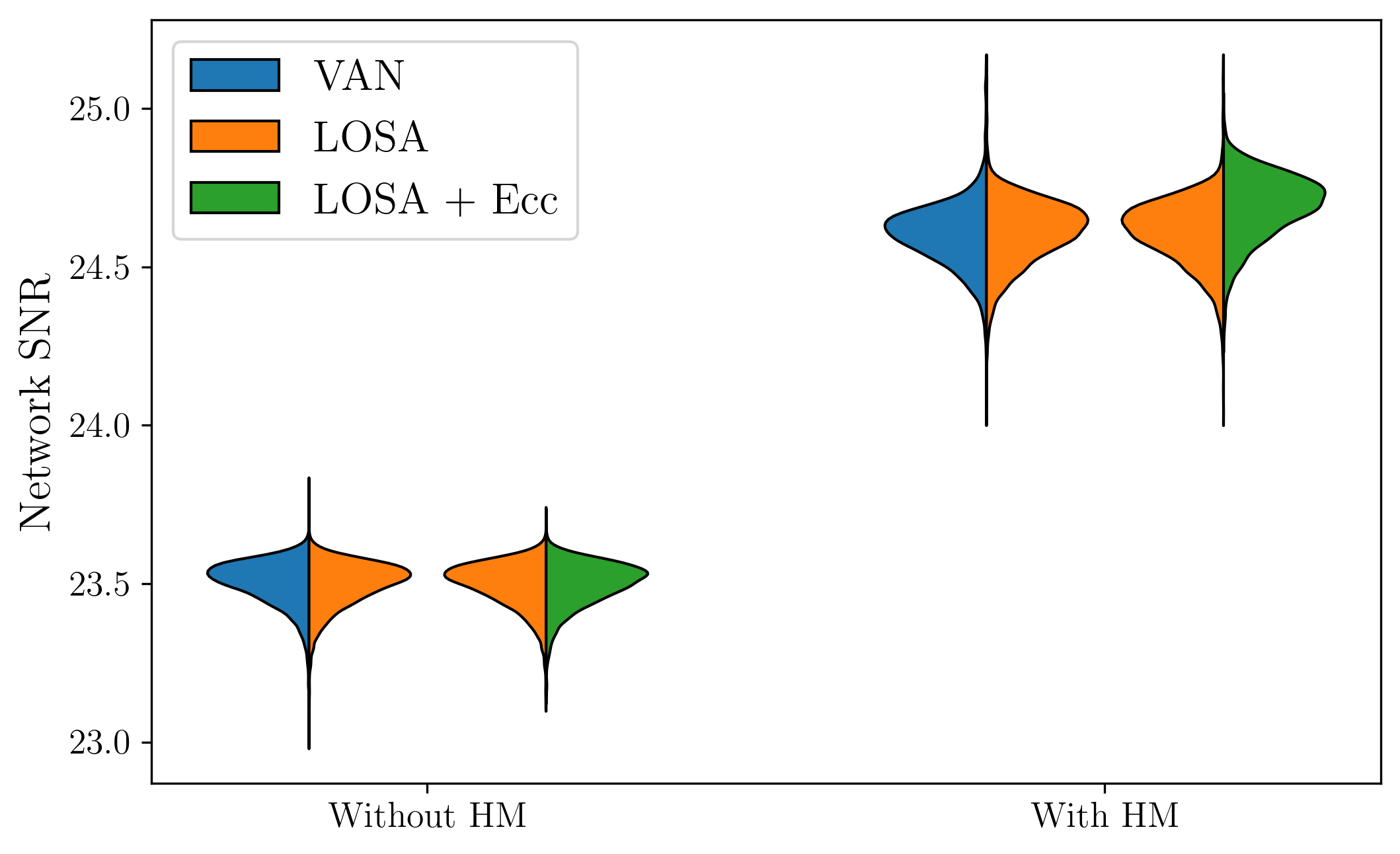}
    \caption{
Posterior distributions of the network matched-filter SNR for GW190814 under three waveform models: a quasi-circular baseline model (VAN), a LOSA-only model, and a LOSA model combined with residual orbital eccentricity $e_0$ (LOSA + Ecc), as defined 
by Equations \ref{eq:htot} and  \ref{eq:dephasing_equation}.
    Results are shown both without and with HOMs. Each violin plot represents the posterior distribution for a given model, with width indicating the relative probability density. Including HOMs increases the recovered SNR for all models. The LOSA-only and LOSA + Ecc models yield SNRs comparable to those of the baseline model, indicating that including these additional effects does not significantly improve the fit to the data.}
    \label{fig:matched_filter_SNR}
\end{figure}

Furthermore, we perform an additional analysis using only the dominant $(2,2)$ mode (i.e., excluding HOMs) to assess the impact of higher modes on LOSA inference. Although the SNR contribution from HOMs is not expected to be large (see Fig.~\ref{fig:matched_filter_SNR}), we find that their absence effectively removes any measurable information about LOSA, with the posterior reverting to the prior.

\subsection{$e_0$-Only Analysis}
\label{subsec:e0_only_analysis}
We perform the $e_0$-only analysis both with and without HOMs. The eccentricity estimates obtained when including HOMs are consistent with previously reported results by~\cite{Kacanja_2025}, who reanalyzed GW190814 using the \texttt{SEOBNRv5EHM}~\citep{PhysRevD.108.124035,PhysRevD.108.124038,jxrc-z298,mihaylov2023pyseobnrsoftwarepackagegeneration} and \texttt{TEOBResumS-Dali}~\citep{PhysRevD.110.084001} waveform models. These models describe aligned-spin BBH systems and incorporate both HOMs and orbital eccentricity.
In contrast, restricting the analysis to the dominant $(2,2)$ mode yields uninformative posteriors for $e_0$, underscoring the importance of HOMs for such MCB systems. For completeness, we also perform an analysis using a $4$ seconds signal duration and find a small but marginally non-zero estimate of the eccentricity ${(e_0 \sim 0.02)}$. However, this result should be treated with caution due to the insufficient signal duration, as discussed earlier.
\begin{figure*}[!hbt]
    \centering
    \includegraphics[width=0.95\linewidth]{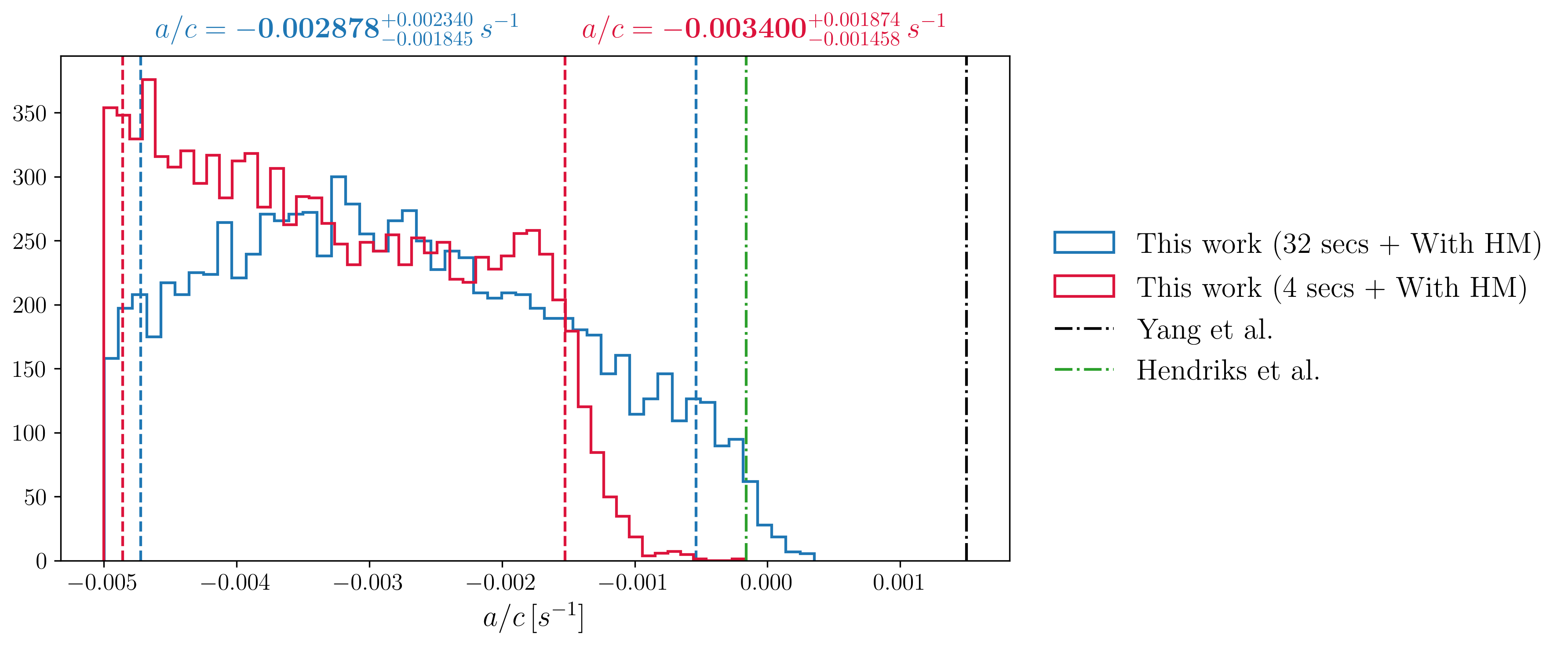}
    \caption{
    Posterior distributions of the LOSA parameter ($a/c$) obtained from a joint analysis for GW190814 that incorporates dominant-order LOSA and residual eccentricity effects into the \texttt{IMRPhenomXPHM} waveform family, while employing two different signal durations. The blue histogram corresponds to a $32$-second analysis, while the red histogram shows the results for a $4$-second analysis (vertical dashed lines indicate the $90\%$ credible intervals for each case). The green dashed-dotted line marks the result reported by~\cite{hendriks2026_LOSA}, while the black dash-dotted line denotes the value inferred by~\cite{Yang2025_LOSA}. Unlike the LOSA-only analysis, the joint analysis yields non-zero estimates of $a/c$ for both signal durations, likely induced by the strong correlation between LOSA and residual eccentricity effects. However, the broad and asymmetric posteriors, together with the low statistical significance of the joint model, indicate that these estimates are driven by degeneracy between the two effects rather than robust evidence for LOSA. Furthermore, the results from the $4$-second analysis should be treated with caution due to the limited signal duration. Unless stated otherwise, our baseline waveform model incorporates HOMs, as implemented in \texttt{IMRPhenomXPHM}~\citep{PhysRevD.103.104056}.
    }
    \label{fig:losa_ecc_both_losa_comparion_32sec_vs_4sec_withHM}
\end{figure*}
\subsection{Joint LOSA and $e_0$ Analysis}
\label{subsec:losa_e0_joint_analysis}
\begin{figure*}[!hbt]
    \centering
    \includegraphics[width=0.95\linewidth]{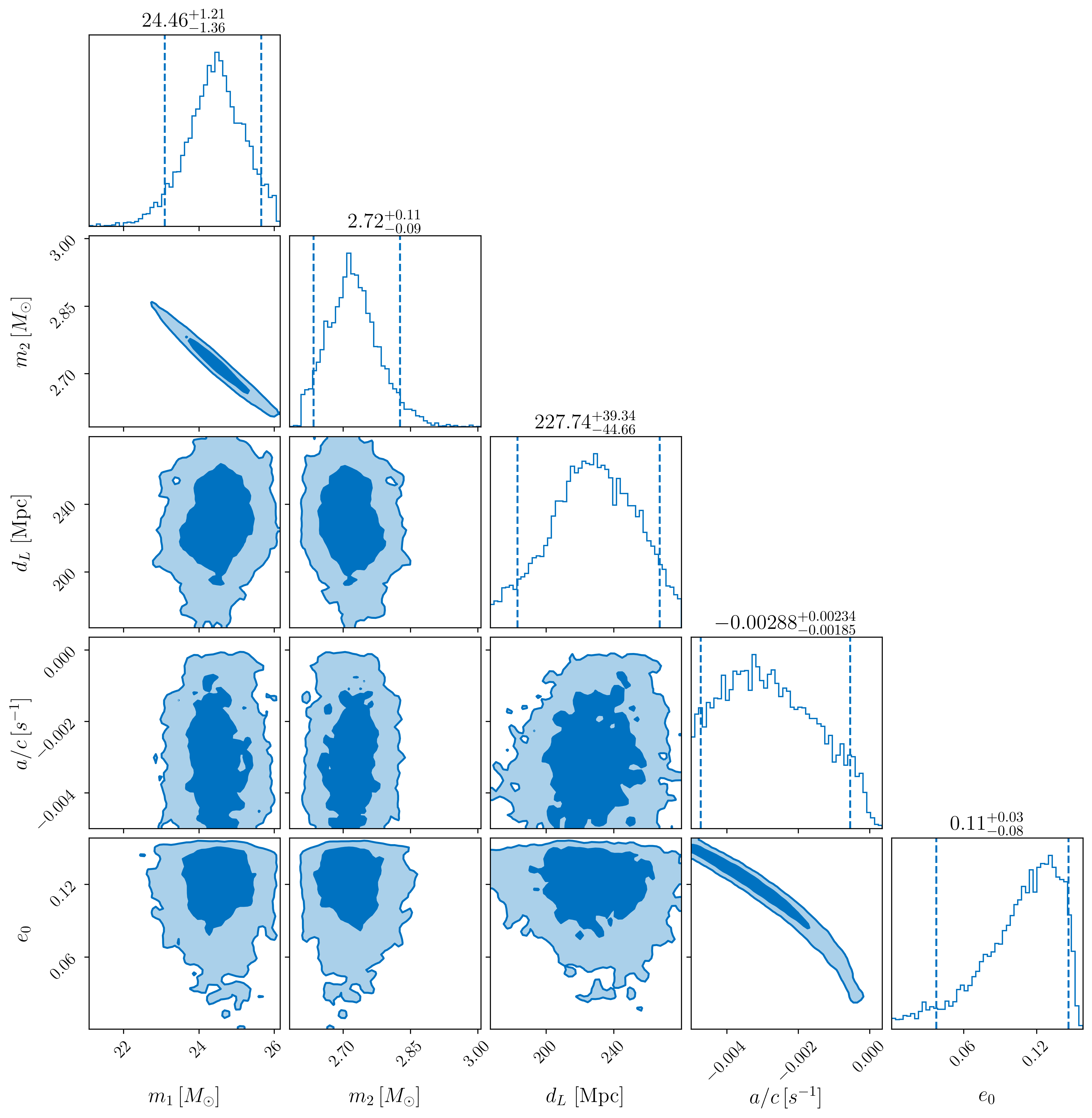}
    \caption{
Posterior distributions for a few selected source parameters of GW190814 obtained from a joint analysis that incorporates dominant-order LOSA and residual eccentricity effects into the \texttt{IMRPhenomXPHM} waveform family, employing $32$ seconds of data. Shown are the component masses ($m_1$, $m_2$), luminosity distance ($d_L$), LOSA parameter $a/c$, and the initial eccentricity $e_0$. The diagonal panels display marginalized one-dimensional posteriors, with dashed vertical lines indicating the $90\%$ credible intervals. The off-diagonal panels show two-dimensional posterior distributions, with contours corresponding to the $50\%$ and $90\%$ credible regions. A strong correlation between $a/c$ and $e_0$ is evident, indicating a degeneracy in their imprint on the inspiral GW phase. This degeneracy most likely explains the non-zero posterior support for both parameters, despite the lack of strong statistical evidence for either effect individually. We use the following PE settings for dynesty sampler as implemented in bilby: \texttt{nlive = 1024}, \texttt{dlogz = 0.1}, \texttt{sample = "acceptance-walk"}, and \texttt{naccept = 60}.
    }
    \label{fig:corner_plot_32secs_a_e_0_joint}
\end{figure*}
We now discuss the results of our PE studies that incorporated both LOSA and $e_0$ effects into our baseline model, as evident from Eq.~\ref{eq:dephasing_equation}. This joint analysis allows us to probe potential degeneracies between the two effects and assess whether their simultaneous inclusion alters the inferred posterior distributions compared to the individual analyses presented above.
As in the LOSA-only analysis, we perform both all-mode and $(2,2)$-only analyses. Unlike the LOSA-only case, where the posterior for $a$ shows no support for LOSA, the joint analysis with all modes included yields informative posteriors and non-zero estimates for both LOSA (${a/c \sim  -2.8 \times 10^{-3} \mathrm{s}^{-1}}$) and eccentricity (${e_0 \sim 0.11}$), with a clear correlation between them. The Bayes factor comparing this run to the \emph{vanilla} run is ${\mathrm{BF^{LOSA+Ecc}_{VAN}} \sim 0.64}$ ($\Delta \log_{10} \mathrm{BF^{LOSA + Ecc}_{VAN}} \sim -0.19$), indicating a modest increase relative to the LOSA-only analysis, but not sufficient to claim evidence for or against the presence of LOSA and eccentricity. Performing the same analysis with a signal duration of $4$ seconds yields similar inferred values for both parameters, namely ${a/c \sim  -3.4 \times 10^{-3} \mathrm{s}^{-1}}$ and ${e_0 \sim 0.09}$ with a corresponding $\mathrm{BF^{LOSA \,+\, Ecc}_{VAN}} \sim 1104$ ($\Delta \log_{10} \mathrm{BF^{LOSA + Ecc}_{VAN}} \sim 3.04$), while retaining the correlation between $a/c$ and $e_0$. However, we do not consider a $ 4$-second signal duration to be appropriate for robust inference and therefore treat these results with caution. In contrast, when the analysis is restricted to the dominant $(2,2)$ mode, the posteriors for both $a/c$ and $e_0$ revert to the priors, indicating the importance of HOMs. 
\section{conclusion}
\label{sec:conclusion}
The leading-order contributions of LOSA and $e_0$ to the Fourier-domain phase of the inspiral waveform scale as $f^{-13/3}$ and $f^{-34/9}$, respectively. This similarity in frequency dependence implies that a GW phase modulation attributed to LOSA could, in principle, be mimicked by a small eccentricity effect, and vice versa. Consequently, these effects are expected to exhibit a strong correlation, which we demonstrate using match estimates. Specifically, we compare an IMR waveform family that incorporates both LOSA and $e_0$ effects against a baseline IMR model that neglects these additional astrophysical contributions. The resulting match analysis reveals a clear degeneracy, motivating a joint reanalysis of the GW190814 event, for which a non-zero LOSA was previously reported by~\cite{Yang2025_LOSA}.

We reanalyze GW190814 by performing PE studies with LOSA and orbital eccentricity included both individually and jointly, using a modified \texttt{IMRPhenomXPHM} waveform family
while typically employing $32$ seconds of data.
In the LOSA-only analysis, we find no evidence for a non-zero LOSA, consistent with the non-detections reported by~\cite{hendriks2026_LOSA} and with the suggestion by~\cite{tiwari2025b} that the reported LOSA signature may arise from waveform systematics, while being in tension with the claim by~\cite{Yang2025_LOSA}. For completeness, we repeat the LOSA-only analysis using a $4$ seconds data segment and recover a non-zero estimate of LOSA (${a/c \sim -1.4 \times 10^{-3}\,\mathrm{s}^{-1}}$) with high statistical significance (as quantified by ${\mathrm{BF}^{\mathrm{LOSA}}_{\mathrm{VAN}} \sim 567}$, where VAN denotes the baseline waveform without LOSA). However, such inferences should be treated with caution due to the insufficient signal duration, as emphasized by~\cite{hendriks2026_LOSA}.

In the joint analysis including both LOSA and eccentricity, we obtain non-zero point estimates for both parameters, ${a/c \sim -2.8 \times 10^{-3}\,\mathrm{s}^{-1}}$ and $e_0 \sim 0.11$, with a strong degeneracy between them, as anticipated from the similarity in their frequency dependence discussed earlier. However, the statistical significance of these estimates is low, as indicated by the Bayes factor ${\mathrm{BF}^{\mathrm{LOSA+Ecc}}_{\mathrm{VAN}} \sim 0.64}$, and the data do not provide compelling evidence for the presence of either effect. We therefore interpret these non-zero estimates as being driven by the degeneracy between LOSA and eccentricity, rather than as robust evidence for LOSA. Restricting the waveform model to the dominant $(2,2)$ mode yields uninformative (flat) posteriors for both $a/c$ and $e_0$, highlighting the importance of HOMs, even though their individual contribution to the SNR is modest.

We also perform eccentricity-only analyses using a $32$ seconds signal duration, both with and without HOMs, as well as a $4$ seconds analysis with HOMs included. Consistent with the joint analysis, removing HOMs yields uninformative posteriors for $e_0$, whereas including them yields estimates broadly consistent with previous studies~\citep{Kacanja_2025}. In contrast, the $4$ seconds analysis yields a marginal but non-zero estimate of eccentricity ($e_0 \sim 0.02$), which should again be treated with caution due to the insufficient signal duration.

Influenced by the above inferences, we are pursuing several investigations that explore possible ways to uniquely determine the presence of LOSA and/or $e_0$ effects in transient 
GW events. A natural direction is to incorporate higher-order LOSA and $e_0$ effects, as provided in \cite{Vijaykumar_2023,hemanta2025eccentricity}. However, our parameter estimation runs that include $\mathcal{O}(e_0^4)$ contributions, along with a recent effort by~\cite{Vijaykumar_2023}, suggest that such additions will not be relevant for GW190814. We note that \cite{Vijaykumar_2023} analyzed a GW170608-like system~\citep{Abbott_2017,PhysRevX.9.031040} with a total mass of $\sim 18.5\, M_{\odot}$ and found no significant bias when restricting LOSA dephasing to leading order. Since GW190814 has a higher total mass ($\sim 25.9\,M_{\odot}$), the impact of higher-order PN corrections is expected to be even smaller, given the shorter inspiral interval within the observatory's frequency band.

We note that HOMs play a nontrivial role in constraining LOSA and eccentricity in GW190814. While their contribution to the total network SNR (see Fig.~\ref{fig:matched_filter_SNR}) is modest, analyses excluding HOMs yield largely uninformative LOSA posteriors, whereas HOMs-inclusive analyses favour posteriors consistent with zero LOSA. In the joint LOSA–eccentricity analysis, we observe a similar pattern in the analysis without HOMs, although substantial correlation persists. This suggests that although the dominant inspiral phasing carries the primary LOSA and eccentricity signatures, HOMs provide additional waveform structure that improves the analysis's constraining power by helping break degeneracies with the intrinsic binary parameters. Our implementation introduces LOSA and eccentricity through leading-order frequency-dependent phase corrections applied to the full IMRPhenomXPHM waveform. This phenomenological treatment captures the dominant inspiral dephasing associated with LOSA and residual eccentricity, but does not include the mode-dependent amplitude and phase corrections expected in fully self-consistent eccentric HOMs waveforms. 
It is possible to develop such an eccentric inspiral $\tilde h(f)$ that incorporates the effects of orbital, periastron and GW emission effects for spinning compact binaries with inputs from~\cite{Tiwari_2019, Henry_2023}.

More promising opportunities may arise during the era of third-generation (3G) GW observatories, such as Cosmic Explorer~\citep{reitze2019cosmicexploreruscontribution,Hall:2022dik} and the Einstein Telescope~\citep{Punturo:2010zz}, owing to their significantly improved low-frequency sensitivity. Since both LOSA and orbital eccentricity predominantly affect the low-frequency regime of the signal, incorporating these effects into IMR waveform models will be essential for 3G data analysis. For such observatories, it will be critical to include higher-order PN contributions associated with these effects, available in \cite{Vijaykumar_2023,hemanta2025eccentricity}. These may even play a decisive role in breaking degeneracies between LOSA and eccentricity effects in transient GW events during the 3G era. 
Similar efforts may also be relevant for IMR events observed by 
LISA~\citep{amaroseoane2017laserinterferometerspaceantenna, Tiwari_2023}, Taiji~\citep{Hu:2017mde}, TianQuin~\citep{Luo_2016}, DECIGO~\citep{Isoyama_2018, kawamura2020currentstatusspacegravitational, Tiwari_2023}, IndiGO-D~\citep{sharma2026indigodprobingcompactbinary} and other planned and proposed space missions.

\begin{acknowledgments}
We thank Avinash Tiwari for a thorough reading of the manuscript during the LVK's internal Publications and Presentations (P\&P) review and for his insightful suggestions, which significantly improved its clarity. L.~P.\ is supported by the postdoctoral fellowship offered by the Inter-University Centre for Astronomy and Astrophysics (IUCAA), Pune. P.~H.\ gratefully acknowledges the support of the Department of Physics at The Chinese University of Hong Kong through the Postgraduate Studentship that made this research possible. P.~H.\ also acknowledges support from the Research Grants Council of Hong Kong (Project Nos.\ CUHK~14304622 and 14307923), the start-up grant from The Chinese University of Hong Kong, and the Direct Grant for Research from the Research Committee of The Chinese University of Hong Kong. A.~G.\ acknowledges the support of the Department of Atomic Energy, Government of India, under project identification RTI~4002, 
and the CAS President’s International Fellowship Initiative (2026PVA0020).

The authors are grateful for the computational resources provided by the LIGO Laboratory and supported by the National Science Foundation Grants No. PHY-0757058 and No. PHY-0823459. This material is based upon work supported by NSF's LIGO Laboratory, which is a major facility fully funded by the National Science Foundation. We are grateful for the computational resources provided by the Leonard E Parker
Centre for Gravitation, Cosmology and Astrophysics at the University of Wisconsin-Milwaukee. This research has made use of data or software obtained from the Gravitational Wave Open Science Center~\citep{Abbott_2023}, a service of the LIGO Scientific Collaboration, the Virgo Collaboration, and KAGRA. This material is based upon work supported by NSF's LIGO Laboratory, which is a major facility fully funded by the National Science Foundation, as well as the Science and Technology Facilities Council (STFC) of the United Kingdom, the Max-Planck-Society (MPS), and the State of Niedersachsen/Germany for support of the construction of Advanced LIGO and construction and operation of the GEO600 detector. Additional support for Advanced LIGO was provided by the Australian Research Council. Virgo is funded through the European Gravitational Observatory (EGO), the French Centre National de Recherche Scientifique (CNRS), the Italian Istituto Nazionale di Fisica Nucleare (INFN), and the Dutch Nikhef, with contributions by institutions from Belgium, Germany, Greece, Hungary, Ireland, Japan, Monaco, Poland, Portugal, Spain. KAGRA is supported by the Ministry of Education, Culture, Sports, Science and Technology (MEXT), Japan Society for the Promotion of Science (JSPS) in Japan; National Research Foundation (NRF) and the Ministry of Science and ICT (MSIT) in Korea; Academia Sinica (AS) and National Science and Technology Council (NSTC) in Taiwan.
\end{acknowledgments}

\appendix
\section{Injection Studies}
\label{sec:appendix}
We performed several zero-noise injection studies using IMRPhenomXPHM as the base waveform model. These studies are designed to assess the
recoverability of LOSA and eccentricity parameters, investigate their mutual
degeneracy, and understand the role of HOMs in the
inference. Unless otherwise stated, the injections are generated using the standard IMRPhenomXPHM waveform without LOSA or eccentricity corrections. For all PE runs in this appendix, we employ the
\texttt{Nessai}~\citep{Williams_2021} sampler (a normalizing-flow-based nested
sampler~\citep{Kobyzev_2021,papamakarios2021normalizingflowsprobabilisticmodeling})
as implemented in \texttt{bilby}, with
\texttt{nlive = 1024},
\texttt{dlogz = 0.1},
\texttt{flow\_class = "GWFlowProposal"},
and \texttt{analytic\_priors = True}. We adopt the same priors as listed in Table~\ref{tab:prior_table}, except for the luminosity distance, for which we use a broad prior range of $10$-$10^4$ Mpc.

First, we inject a non-accelerating and non-eccentric GW190814-like signal into zero noise and recover it using the modified IMRPhenomXPHM waveform model with LOSA corrections, considering both HOMs-inclusive and dominant $(2,2)$-mode-only analyses. The injection parameters are set to:
${m_1 = 23\,M_{\odot}}$,
${m_2 = 2.6\,M_{\odot}}$,
${a_1 = 0.04}$,
${a_2 = 0.57}$,
${\theta_1 = 1.94\,\mathrm{rad}}$,
${\theta_2 = 1.46\,\mathrm{rad}}$,
${d_L = 257.5\,\mathrm{Mpc}}$,
${\theta_{jn} = 0.85\,\mathrm{rad}}$,
${\phi = 2.91\,\mathrm{rad}}$,
${t_c = 1249852257.012944\,\mathrm{s}}$,
${\alpha = 0.22\,\mathrm{rad}}$,
${\delta = -0.43\,\mathrm{rad}}$,
and ${\psi = 0.40\,\mathrm{rad}}$. In both cases, we recover the injected values of the chirp mass ($\mathcal{M}$), mass ratio ($q$), and LOSA parameter ($a/c$), although the $(2,2)$-only analysis yields broader posteriors.\footnote{
The luminosity distance is recovered accurately in the HOMs-inclusive analysis.
} Next, we recover the same injection using the modified IMRPhenomXPHM waveform model, including both LOSA and eccentricity corrections. In both the HOMs-inclusive and dominant $(2,2)$-mode-only analyses, we observe an apparent offset toward finite eccentricity, accompanied by a correlated shift in LOSA. The effect is substantially larger in the $(2,2)$-only analysis, while the inclusion of HOMs significantly improves the recovery and reduces the extent of the bias. This behaviour is likely driven by the strong LOSA-eccentricity degeneracy together with prior-boundary effects near $e_0 = 0$.

Finally, we inject an accelerating and eccentric GW190814-like signal with $a/c \sim -0.0015\,\mathrm{s}^{-1}$ and $e_0 = 0.05$, and recover it using the modified IMRPhenomXPHM waveform model including both LOSA and eccentricity corrections, with and without HOMs. In both cases, we successfully recover the injected chirp mass, mass ratio, LOSA, and eccentricity parameters.

We note that the present implementation does not incorporate fully self-consistent mode-dependent LOSA and eccentricity corrections in the higher harmonics. Rather, the leading-order inspiral dephasing is applied phenomenologically to the full IMRPhenomXPHM waveform. Therefore, the injection and recovery studies presented here should primarily be interpreted as consistency checks of the leading-order implementation in an idealized zero-noise setting. In particular, successful recovery of injected LOSA and eccentricity parameters in these studies should not be interpreted as evidence that the LOSA--eccentricity degeneracy or mimicry is fully resolved. A more realistic injection study would require waveform models incorporating higher-order PN contributions together with mode-dependent corrections to the higher harmonics, which is beyond the scope of the present work.

\begin{figure*}[p]
\centering

\subfloat[
Recovery of a non-accelerating and non-eccentric injection using a LOSA-only
recovery model.
\label{fig:inj_van_rec_losa}
]{
\includegraphics[width=0.47\textwidth]
{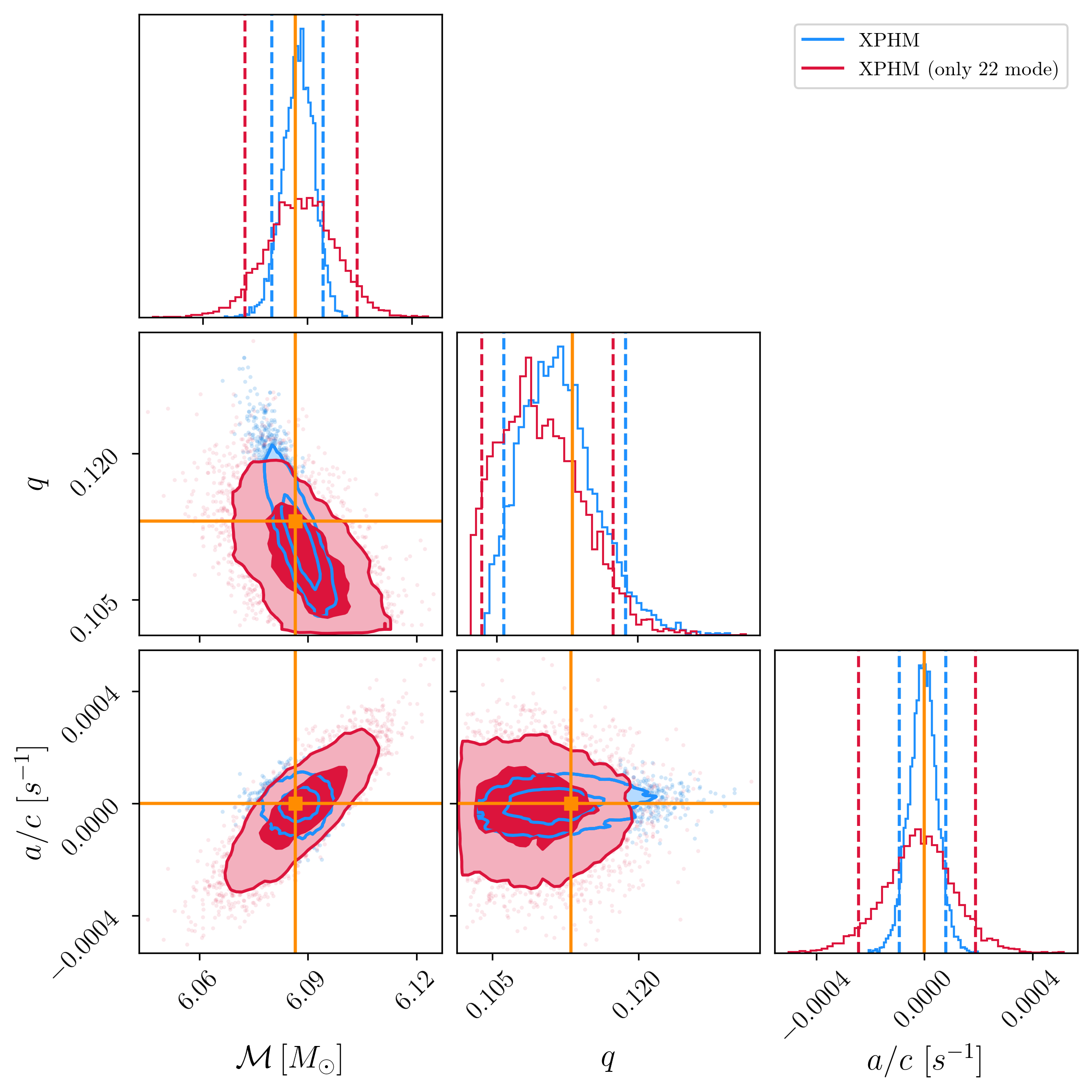}
}
\hfill
\subfloat[
Recovery of a non-accelerating and non-eccentric injection using a joint LOSA
and eccentricity recovery model.
\label{fig:inj_van_rec_losa_e0}
]{
\includegraphics[width=0.47\textwidth]
{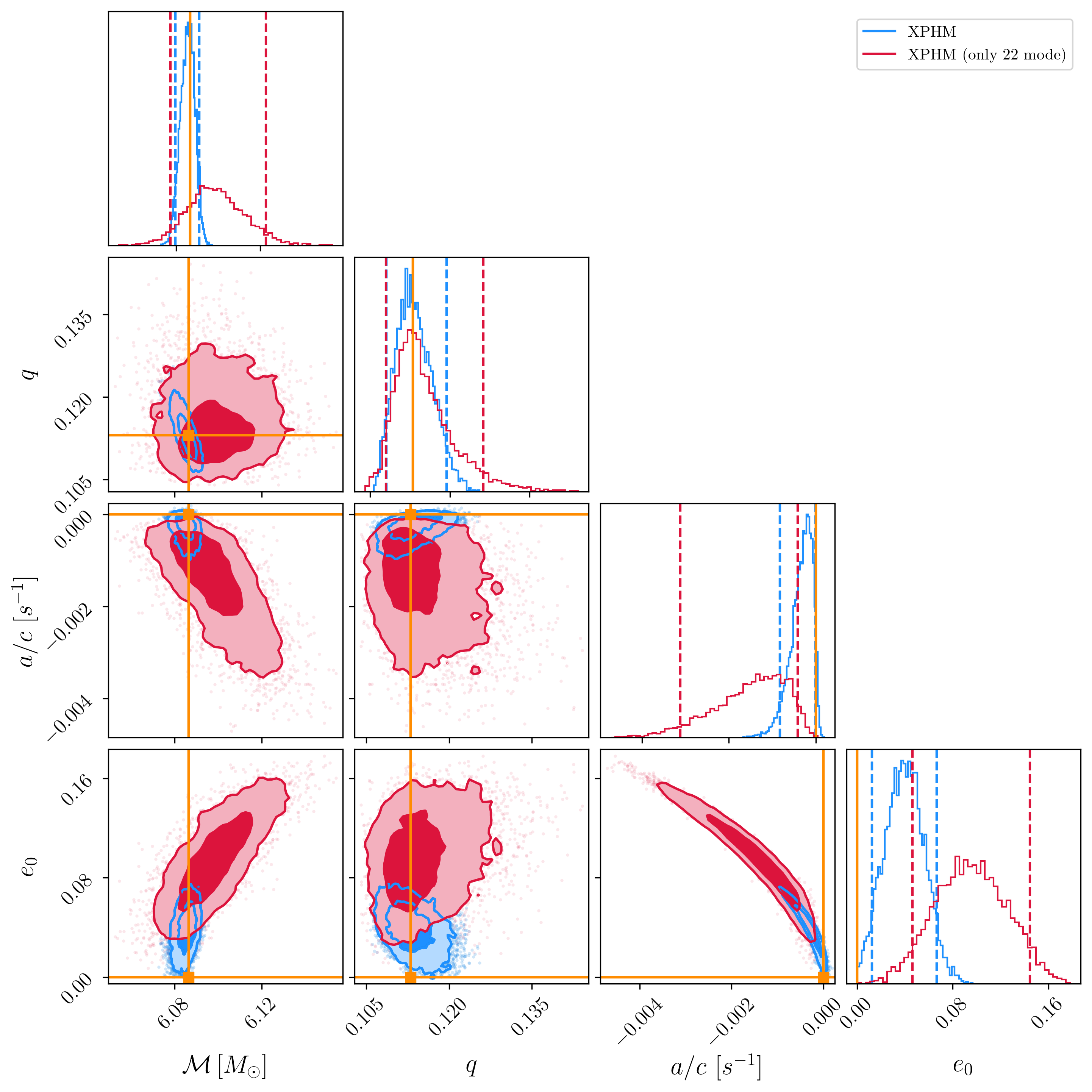}
}

\vspace{0.4cm}

\subfloat[
Recovery of an accelerating and eccentric injection with ${a/c \sim -0.0015\,\mathrm{s}^{-1}}$ and ${e_0 = 0.05}$ using a joint LOSA and
eccentricity recovery model.
\label{fig:inj_losa_e0_rec}
]{
\includegraphics[width=0.60\textwidth]
{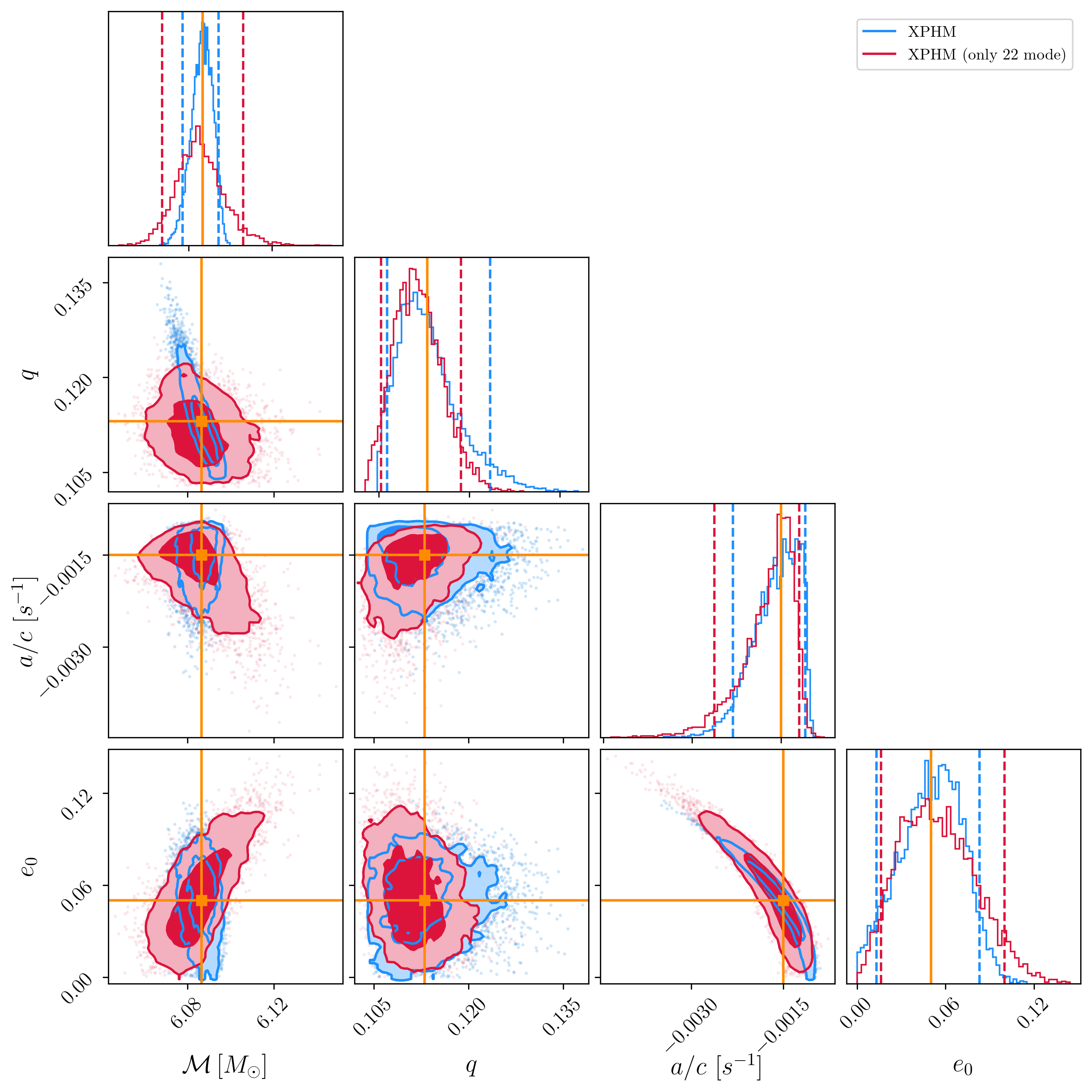}
}

\caption{
Zero-noise injection studies for GW190814-like signals using modified
\texttt{IMRPhenomXPHM} waveform models including LOSA and eccentricity
corrections. Blue contours correspond to analyses including HOMs, while red contours correspond to analyses restricted to the dominant
$(2,2)$ mode. Panel (a) shows the recovery of a quasi-circular and non-accelerating
injection using a LOSA-only recovery model. Panel (b) shows the recovery of
the same injection using a recovery model that includes both LOSA and
eccentricity. Panel (c) shows the recovery of an injection with both LOSA and
eccentricity present in the signal. The results illustrate the strong LOSA-eccentricity degeneracy discussed in the main text and demonstrate that including HOMs improves the recovery of these parameters and reduces the extent of degeneracy-induced biases.
}
\label{fig:appendix_injections}
\end{figure*}

\bibliographystyle{aasjournal}
\bibliography{references}

\end{document}